\documentclass[aps,superscriptaddress,showpacs,showkeys]{revtex4}

\usepackage[final]{graphicx}
\usepackage{t1enc}

\begin{document}

\title{Testing boundary conditions efficiency in simulations of long-range interacting magnetic models}

\author{Sergio A. Cannas}
\email{cannas@famaf.unc.edu.ar}
\affiliation{Facultad de  Matem\'atica, Astronom\'{\i}a  y F\'{\i}sica, Universidad 
Nacional de C\'ordoba, \\ Ciudad Universitaria, 5000 C\'ordoba, Argentina}
\altaffiliation{Member of CONICET, Argentina}
\author{Cintia M. Lapilli}
\email{cmlff5@mizzou.edu}
\affiliation{Department of Physics and Astronomy, University of Missouri-Columbia.
Columbia, Missouri 65211, USA}
\author{Daniel A. Stariolo}
\email{stariolo@if.ufrgs.br}
\affiliation{Departamento de F\'{\i}sica\\
Universidade Federal do Rio Grande do Sul\\
CP 15051, 91501--979, Porto Alegre, Brazil}
\altaffiliation{Research Associate of the Abdus Salam International
Centre for Theoretical Physics (ICTP), Trieste, Italy}

\date{\today}

\begin{abstract}
Periodic boundary conditions have not a unique implementation in magnetic systems where all spins interact with each other through a power law decaying interaction of the form $1/r^\alpha$, $r$ being the distance between spins.  In this work we present a comparative study of the finite size effects oberved in numerical simulations by using first image convention and full infinite of periodic boundary conditions in one and two-dimensional spin systems with those type of interactions, including the ferromagnetic, antiferromagnetic and competitive interactions cases. Our results show no significative differences  between the finite size effects produced by both types of boundary conditions when the low temperature phase has zero global magnetization, while it  depends on the ratio $\alpha/d$ for systems with a low temperature ferromagnetic phase. In the last case the first image convention gives much more stronger finite size effects than the other when the system enters into the classical regime $\alpha/d \leq 3/2$.
\end{abstract}

\maketitle

\section{Introduction}

Boundary conditions are a central issue in almost every statistical mechanics calculation. In particular, 
periodic boundary conditions are the usual way of diminishing finite size border effects in translationally  invariant  systems. This point is of special importance in numerical simulations, where an appropriated handling of finite size effects may determine completely  the quality of the results. While both the interpretation and implementation of periodic boundary conditions are straigthforward in systems with short range microscopic interactions, their usage in systems with long-range interactions is not so easy, with several subtleties to take into account. In this work we will consider magnetic systems where the iteractions between pairs of spins decay as $1/r^\alpha$, $r$ being the distance between spins and $\alpha>0$ (this includes, among the most conspicuous examples, Coulomb and dipolar interactions). Then, every spin interacts with any other spin in the system, with a slow decaying distance-dependent intensity. Finite size effects are much more stronger in these kind of systems than in systems with short range interactions, because  every spin feels {\it directly} the influence of the border. Hence, boundary conditions are a topic of central importance in numerical simulations of these kind of systems.

Spin systems with the above type of interactions may present a very rich phenomenology, even in one dimension, and have attracted a lot of attention in the last years, both because its applications to ultrathin magnetic films systems\cite{DeBell}-\cite{Gleiser}, glassy systems \cite{Grousson} and to the study of statistical mechanics fundamentals\cite{Cannas1}-\cite{Tamarit}, among many others.

There are two alternative representations of a finite system with periodic boundary conditions.  Let us think on a spin system defined on a hypercubic $d$-dimensional lattice with $N=L^d$ sites. On one hand, we can visualize the system as a d-torus (closed ring in $d=1$; torus in $d=2$, etc.), where a given spin interacts with its closer neighbors defined by the topology of the d-torus, up  to a certain range of interaction. On the other hand, we can think that we have an infinite system, where  the original finite system has been replicated infinite times in all the coordinate directions. In other words, we can think that the infinite lattice has been partitioned into cells of size $N=L^d$ and that we have chosen only the periodic solutions of the problem, with periodicity $L$ in all the coordinate directions.

While the difference between both views is just a matter of interpretation  for systems with short-range interactions, the situation changes for long-range interacting systems. According to the second scheme, a given spin will interact with {\em infinite replicas of everyone of the rest of the spins} and we can express the effective interaction between two spins inside the system as an infinite sum over replicas (this includes a ``self-interaction'' term, that accounts for the interaction of every spin with its own replicas). We will call these {\it infinite periodic boundary conditions} (IPBC). On the other hand, in the closed topology of the first scheme it is consistent to consider that every pair of spins inside the system interacts only through its {\it minimal} distance over the hypertorus. This corresponds to the minimal truncation of the infinite series of the previous scheme and it is sometimes called the {\it first image convention}; we will refer to it as {\it first image periodic boundary conditions} (FIPBC). While both type of boundary conditions give the same result in the thermodynamic limit $L\rightarrow\infty$, they are clearly different for finite systems. 

Now, the question is how good are both type of boundary conditions to wipe out finite size effects in systems of moderate sizes, such as those available in practical numerical calculations? This is of particular interest,  considering that the numerical simulations are much more costly than in the case of short range interactions (the computational complexity of the algorithms is typically ${\cal O}(N^2)$). Although both IPBC  and FIPBC  have been used in several works, including both numerical\cite{DeBell}-\cite{Cannas1}\cite{Luijten} and analytical\cite{Lee,Campa} calculations, to the best of our knowledge the only comparison of its relative efficiency in magnetic systems was reported in an old work by Kretschmer and Binder \cite{Kretschmer}, in the case of dipolar interactions in a three-dimensional system. In that work the authors observed a very poor performance of the FIPBC compared to the IPBC when the system exhibits a ferromagnetic phase. Since then, several authors assumed this result as a general rule, independent of the dimension $d$, the range of the interactions (i.e., of the exponent $\alpha$) and the symmetry of the low temperature phase. In this work we compare the relative efficiency of both type of boundary conditions by considering the equilibrium critical behaviour of ferromagnetic, antiferromagnetic and competing interactions in $d=1$ and $d=2$ and for different values of the exponent $\alpha$, using a single spin-flip Monte Carlo algorithm (Metropolis).

While the implementation of the FIPBC is straightforward, the implementation of the IPBC is much more cumbersome and needs care, because the infinite series for the effective interactions are usually slowly convergent. This is another reason for studying its relative efficiency. The usual way of handling numerically the IPBC in $d=2$ and $d=3$ is to adapt the Ewald sums technique, originally derived for systems of interacting charged particles\cite{Allen,Frenkel},  to magnetic models\cite{DeBell,Kretschmer}. In section \ref{modeld1} we compare the critical temperature of one dimensional Ising models with ferromagnetic, antiferromagnetic and competing interactions, for different system sizes and $1<\alpha<2$ using FIPBC and IPBC.  In this case we introduce a numerical technique for implementing the IPBC that is  easier to implement in $d=1$ for arbitrary values of $\alpha$ than the Ewald sums technique. We also analized the critical behaviour of the antiferromagnetic model for $0\leq\alpha\leq 1$. In section \ref{modeld2} we compare the specific heat numerical results obtained around the critical region for ferromagnetic long range interactions with $\alpha=3$, and competing ferromagnetic short range interactions with antiferromagnetic (dipolar) long range interactions, using the Ewald sums for the IPBC. A general discussion of the results is presented in section \ref{conclu}.

\section{One-dimensional Ising models}

\label{modeld1}

We considered first an Ising model of the type

\begin{equation}
H= -J_0 \sum_{(i,j)} f(r_{ij}) \; S_i S_j \;\;\;\;\;\; S_i=\pm 1
\label{H1}
\end{equation}

\noindent where  the sum $\sum_{(i,j)}$ runs over all distinct pairs of sites $(i,j)$ ($i,j=1,2,\ldots$) in an infinite chain; $r_{ij}\equiv |i-j|$ is the distance between sites and

\begin{equation}
f(r) = \left\{ \begin{array}{lll}
\frac{1}{r^\alpha}&if& r\geq 1 \\
0 &if& r=0
\end{array} \right.
\label{f}
\end{equation}

\noindent This model presents a phase transition at a non-zero critical temperature $T_c(\alpha)$ for $1<\alpha\leq 2$ \cite{Dyson,Frohlich}. In the ferromagnetic version, $J_0>0$, the critical temperature behaves asymptotically as $k_b T_c/J_0 \sim 2 \zeta(\alpha) \sim 2/(1-\alpha)$ when $\alpha\rightarrow\  1^+$ \cite{Cannas4,Luijten}, where $ \zeta(\alpha) = \sum_{n=0}^\infty \frac{1}{n^\alpha} $ is the Riemann zeta function.
When $1<\alpha \leq 1.5$ the critical exponents are classical. The thermodynamic limit in the ferromagnetic model is not defined when $0\leq \alpha<1$ and the system presents non-extensive behaviour\cite{Cannas1}; however, with an appropriated regularization procedure it can be shown that the system still presents a phase transition where all the thermodynamic functions (not only the critical exponents) obey mean-field behaviour\cite{Cannas2,Lee}. These results generalize to ferromagnetic models in arbitrary dimension $d$ with $\alpha=d+\sigma$: the thermodynamic limit is well defined for $\sigma>0$. The critical exponents are classical for $0<\sigma<d/2$ and assume 
continuously varying values for $d/2\leq \sigma \leq 2$. When $\sigma>min(d,2)$ the system exhibits short range critical behaviour\cite{Ainzenman}.

In this section we considered a finite chain with $1<\alpha\leq 2$; the finite size version of the Hamiltonian (\ref{H1}) is 

\begin{equation}
H= -J_0 \sum_{(i,j)} W(r_{ij}) \; S_i S_j \;\;\;\;\;\; S_i=\pm 1
\end{equation}

\noindent where  now the sum $\sum_{(i,j)}$ runs over all distinct pairs of sites $(i,j)$ ($i,j=1,2,\ldots,L$) in a chain of lenght $L$, $r_{ij}\equiv |i-j|=0,1\ldots L-1$ and $W(r_{ij})$ is the effective interaction between spins $(i,j)$ resulting from the different types of periodic boundary conditions.

The FIPBC effective interaction $W_F$ is then simply defined as

\begin{equation}
 W_F(r) = f\left( {\rm min} [r,|r- L|] \right) 
\end{equation}

\noindent while the IPBC effective interaction $W_I$ is defined as

\begin{equation}
 W_I(r) = \sum_{n=0,\pm 1,\ldots} f(r+nL) = \frac{1}{L^\alpha} \sum_{n=0,\pm 1,\ldots} \frac{1}{|n+r/L|^\alpha} \;\;\;\;\; for\;\;\; r=1,2,\ldots L-1
\label{WI}
\end{equation}

\noindent and

\begin{equation}
 W_I(0) = \sum_{n=\pm 1,\pm 2,\ldots} f(nL) = \frac{1}{L^\alpha} \sum_{n=\pm 1,\pm 2,\ldots} \frac{1}{|n|^\alpha} =2\frac{\zeta(\alpha)}{L^\alpha}
\end{equation}

\noindent where $n=\pm 1,\pm 2,\ldots, \pm \infty$ numerates the infinite replicas of the system $n=0$. Equation (\ref{WI}) is a slowly convergent series and cannot be truncated in a 
straitforward way. In order to obtain a controllable approximation we rewrite   Eq.(\ref{WI}) as:

\begin{eqnarray}
 W_I(r) &=& \frac{1}{r^\alpha}+  \frac{1}{L^\alpha}  \sum_{n=\pm 1,\ldots,\pm\infty} \frac{1}{|n+r/L|^\alpha} \nonumber \\
   &=& \frac{1}{r^\alpha}+  \frac{1}{L^\alpha}\left[ \sum_{n=\pm 1,\ldots,\pm m} \frac{1}{|n+r/L|^\alpha}
+ \sum_{n=m+1}^\infty  \left(  \frac{1}{|n+r/L|^\alpha} +  \frac{1}{|n-r/L|^\alpha}\right) \right] 
\label{WI2}
\end{eqnarray}

\noindent Since $r<L$, for $\alpha>1$ and $m \gg 1$ we can approximate $W_I(r) \approx W_m(r)$ for $r=1,\ldots,L-1$, where

\begin{eqnarray}
 W_m(r) &=&   \frac{1}{r^\alpha}+  \frac{1}{L^\alpha}\left[ \sum_{n=\pm 1,\ldots,\pm m} \frac{1}{|n+r/L|^\alpha}
 + 2  \sum_{n=m+1}^\infty \frac{1}{n^\alpha} \right] \nonumber \\
   &=& \frac{1}{r^\alpha}+ \frac{2\zeta(\alpha)}{L^\alpha} + \frac{1}{L^\alpha} \sum_{n=\pm 1,\ldots,\pm m} \left( \frac{1}{|n+r/L|^\alpha}
 - \frac{1}{|n|^\alpha} \right)
\label{Wm}
\end{eqnarray}

\noindent The finite approximation $W_m(r)$ can be calculated once for ever at the beginning of the simulation.  In order to control the accuracy of the approximation, we can calculate the differences

\begin{equation}
\Delta W_m(r) \equiv W_{m+1}(r)-W_m(r) =  \frac{1}{L^\alpha}\left[ \frac{1}{|m+1+r/L|^\alpha} +\frac{1}{|m+1-r/L|^\alpha} -\frac{2}{(m+1)^\alpha} \right].
\end{equation}

\noindent Noting that

\[ \Delta W_m(r) < \Delta W_m(L)=\frac{1}{L^\alpha} \left( \frac{1}{(m+2)^\alpha}+\frac{2}{m^\alpha}-\frac{2}{(m+1)^\alpha}\right) \]

\noindent we can estimate the error of the approximation by choosing $m$ such that  $\Delta W_m(L)<\epsilon$; since for $m \gg 1$

\[ \Delta W_m(L) = \frac{1}{L^\alpha(m+1)^\alpha} \left[ \frac{\alpha(\alpha+1)}{(m+1)^2} + {\cal O}\left( m^{-4}\right) \right]  \]

\noindent we have

\[ m > \left( \frac{\alpha(\alpha+2)}{\epsilon L^\alpha} \right)^{\frac{1}{\alpha+2}} -1 \].

\noindent In all the simulations we choose $m$ such that $\epsilon<10^{-6}$ for every value of $L$ and $\alpha$. 

\begin{figure}
\begin{center}
\includegraphics[width=14cm,height=10cm,angle=0]{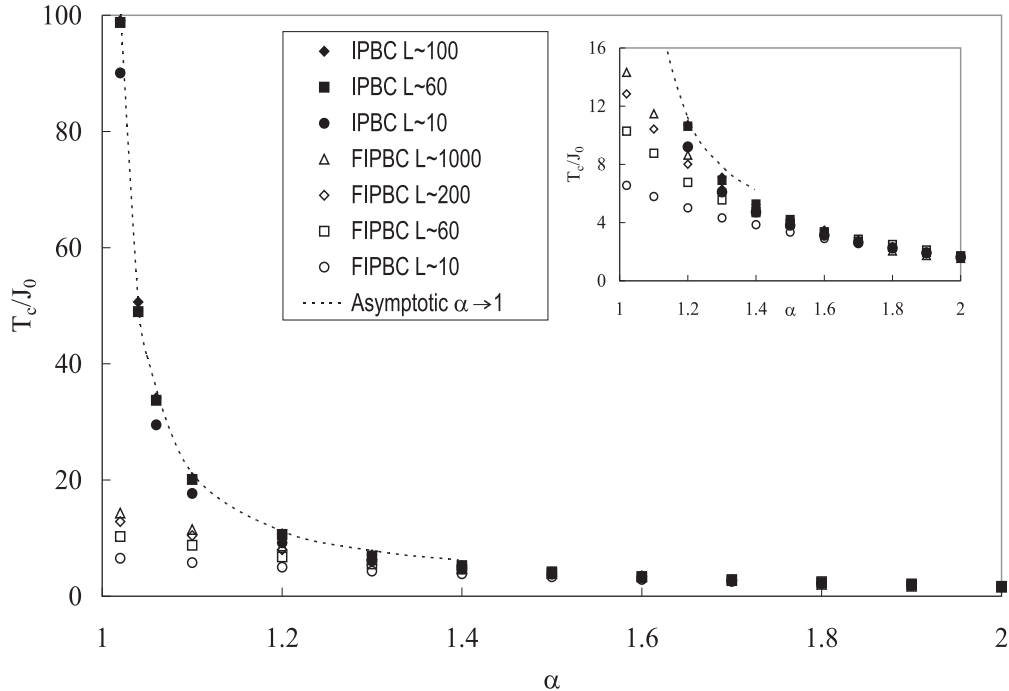}
\caption{ Critical temperature $T_c/J_0$ vs. $\alpha$ ($k_B=1$) for the one-dimensional ferromagnetic model obtained with FIPBC and IPBC for different values of $L$. The notation $L\sim$ indicates the average value of the pairs of values $(L',L")$ used to calculate the Binder cumulant. In the inset it
can be seen the same figure with a magnified vertical scale.}
\label{fig1}
\end{center}
\end{figure}

We first considered the ferromagnetic case $J_0>0$ for $1< \alpha\leq 2$. We calculated the critical temperature using both FIPBC and IPBC by means of the so called Binder cumulant\cite{Binder}

\[ U_L \equiv 1- \frac{\left< M^4 \right>_L}{3\left< M^2 \right>_L^2} \]

\noindent where $M$ is the order parameter (magnetization per spin). The critical temperatures were estimateded as the point of intersection\cite{Binder} of the curves $U_{L'}$, $U_{L"}$ vs. $T$ (in what follows we choose $k_B=1$), for pairs of values $(L',L")$ chosen around some average  $L$, with a  dispersion around $L$ of less than 10\%. The results for boh type of boundary conditions were calculated using the {\it same} single spin-flip Monte Carlo dynamics (Metropolis). The results are shown in Fig.\ref{fig1}, where the exact asymptotic behaviour  when $\alpha\rightarrow 1$ is also shown for comparison.  For $\alpha>1.5$ the results are very similar, with an obtained difference between both types of boundary conditions of less than 5\% and decreasing when $\alpha\rightarrow 2$. On the other hand, for $\alpha<1.5$ the difference in performance increases dramatically as $\alpha$ decreases. In table \ref{table1} the results for the best estimate using IPBC for $\alpha\leq 1.5$ are compared with the best estimates in the literature obtained by Luijten and Bl\"ote\cite{Luijten}, using a specially designed block Monte Carlo algorithm for ferromagnetic systems with long range interactions\cite{Luijten2} (IPBC are implicit in this algorithm) and chains of very large sizes ($L=150000$). While the IPBC gives a rather accurate result, even for a modest size of $L\sim 100$ (see table \ref{table1}), the FIPBC with $L\sim 1000$ give an error that reaches around one order of magnitude for $\alpha=1.02$ (see Fig.\ref{fig1}).

\begin{table}
\begin{tabular}{|l|l|l|}
\hline
$\alpha$ & IPBC $L\sim 100$  & Ref.\cite{Luijten} \\
\hline   
1.1  & $0.049 \pm 0.003$ & $0.0476168$ \\
\hline
1.2  & $0.092 \pm 0.004$  & $0.0922314$ \\
\hline
1.3  &$0.141 \pm 0.005$  & $0.136110$ \\
\hline
1.4  & $0.185 \pm 0.005$  & $0.181150$  \\
\hline
1.5   & $0.238 \pm 0.006$  & $0.229155$  \\
\hline
\end{tabular}  
\caption{\label{table1} Inverse critical temperature $K_c=J_0/T_c$ obtained with $L\sim 100$ IPBC compared with the best estimates from Ref.[9]}
\end{table}

We next considered the antiferromagnetic case $J_0<0$ for $1<\alpha\leq 2$ and system sizes $L=300, 500$ and $1000$ using both type of boundary conditions, and for $0\leq \alpha \leq 1$ using FIPBC (IPBC cannot be implemented in this case). While the ferromagnetic version of this model has been extensively studied,  to the best of our knowledge the only existing results for the antiferromagnetic case for $1<\alpha\leq 2$ have been obtained by Romano (see Ref.\cite{Romano} and references therein). In particular, no previous results are known for $0<\alpha\leq 1$, while the $\alpha=0$ case has been exactly solved\cite{Al-Wahsh}. We found that the system presents a low temperature ordered antiferromagnetic phase for every value of $0<\alpha\leq 1$ below a finite critical temperature. At variance with the ferromagnetic case, the antiferromagnetic model shows a well defined thermodynamic limit for $0<\alpha\leq 1$;  this fact has also been observed in an antiferromagnetic model with long range interactions defined in the hypercubic cell\cite{Franco}.

\begin{figure}
\begin{center}
\includegraphics[width=16cm,height=10cm,angle=0]{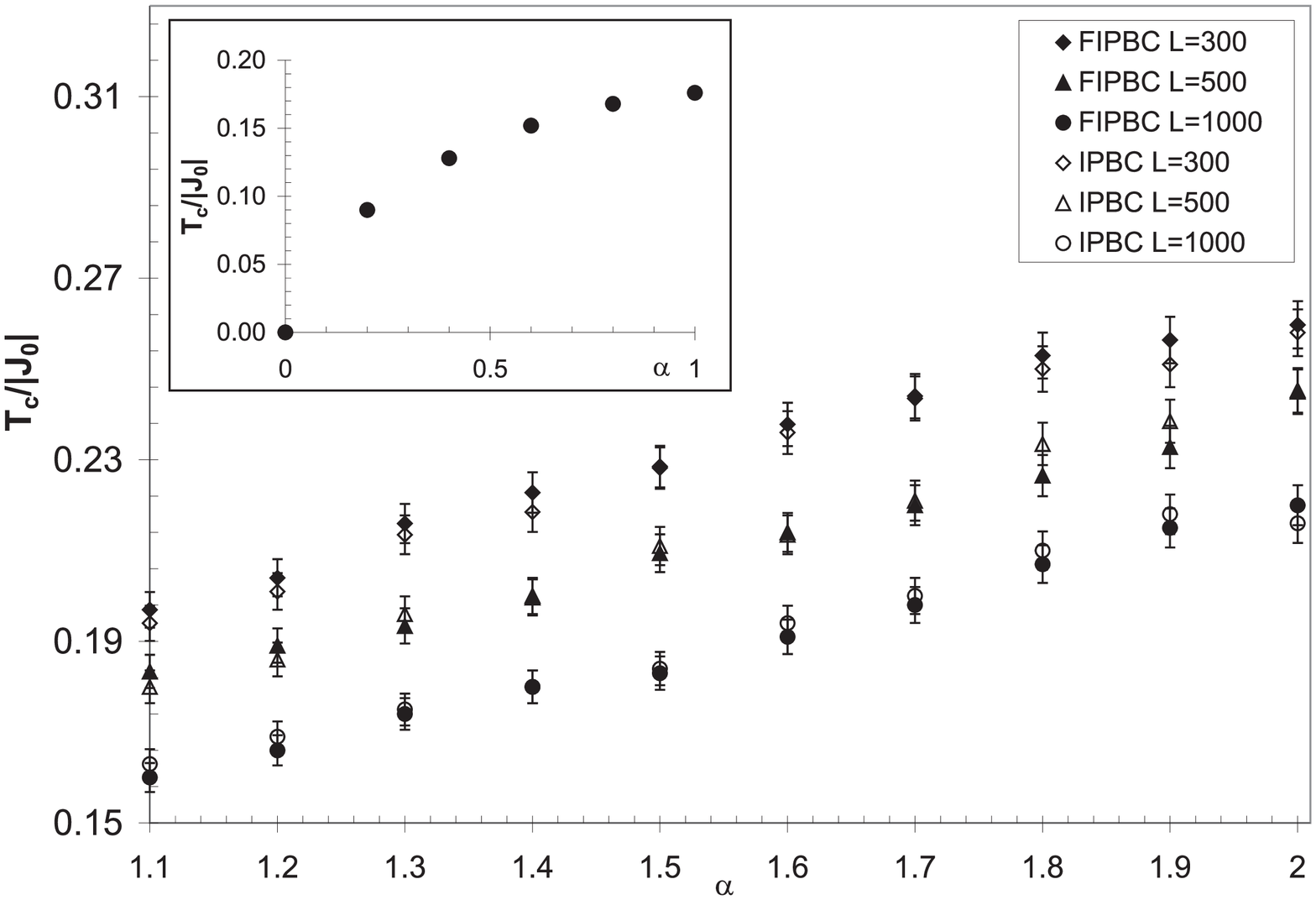}
\caption{Critical temperature $T_c/|J_0|$ vs. $\alpha$ ($k_B=1$) for the one-dimensional antiferromagnetic model obtained with FIPBC and IPBC for different values of $L$.}
\label{fig2} 
\end{center}
\end{figure}

 For this system the critical temperature was estimated as the maximum in the staggered susceptibility, defined as the fluctuations in the staggered magnetization. The results are shown in Fig.\ref{fig2}. We see that both type of boundary conditions give in this case the same result (within the error bars) for every system size when $1<\alpha\leq 2$. The results for $L=1000$ are in good agreement with previous estimations\cite{Romano} using IPBC for $\alpha=2$: $T_c/|J_0| \approx 0.219$, and converge to $T_c =0$ for $\alpha\rightarrow 0$, reproducing the exact solution\cite{Al-Wahsh}.

Finally, we considered an Ising model with competing short range ferromagnetic interactions and long range antiferromagnetic interactions defined by the Hamiltonian

\begin{equation}
H= -J_0 \sum_{<i,j>} S_i S_j  +J_d\sum_{(i,j)} f(r_{ij}) \; S_i S_j \;\;\;\;\;\; S_i=\pm 1,
\label{H2}
\end{equation}

\noindent where $J_0,J_d>0$, the sum $\sum_{<i,j>}$ runs over nearest neighbor paris of spins and $f(r)$ is given by Eq.(\ref{f}). This is a one dimensional generalization of a model for an ultrathin magnetic film with competing ferromagnetic exchange interactions and dipolar interactions that will be described in the next section. The finite temperature phase diagram of this model when $1<\alpha \leq 2$ is similar to the one corresponding to the two dimensional case. At low temperatures it  presents different modulated phases (with zero global magnetization) of period $2h$, depending on the ratio $\delta=J_0/J_d$, with a finite critical temperature $T_c=T_c(\alpha,\delta)$. The period of the modulation increases with $\delta$, ranging from $h=1$ (antiferromagnetic) for low values of $\delta$, up to the system size for very large values of $\delta$, passing by all the possible values commensurable with the system lenght $L$. The calculation of the global phase diagram of this model is in progress and the details will be published elsewhere. 

For the comparison of the boundary conditions we choose some representative values of $\alpha$ and $\delta$; other values gave similar results. The results are shown in table \ref{table2}. No significative differences were observed between both type of boundary conditions.

\section{Two-dimensional Ising models}

\label{modeld2}

\begin{figure}
\begin{center}
\includegraphics[width=16cm,height=10cm,angle=0]{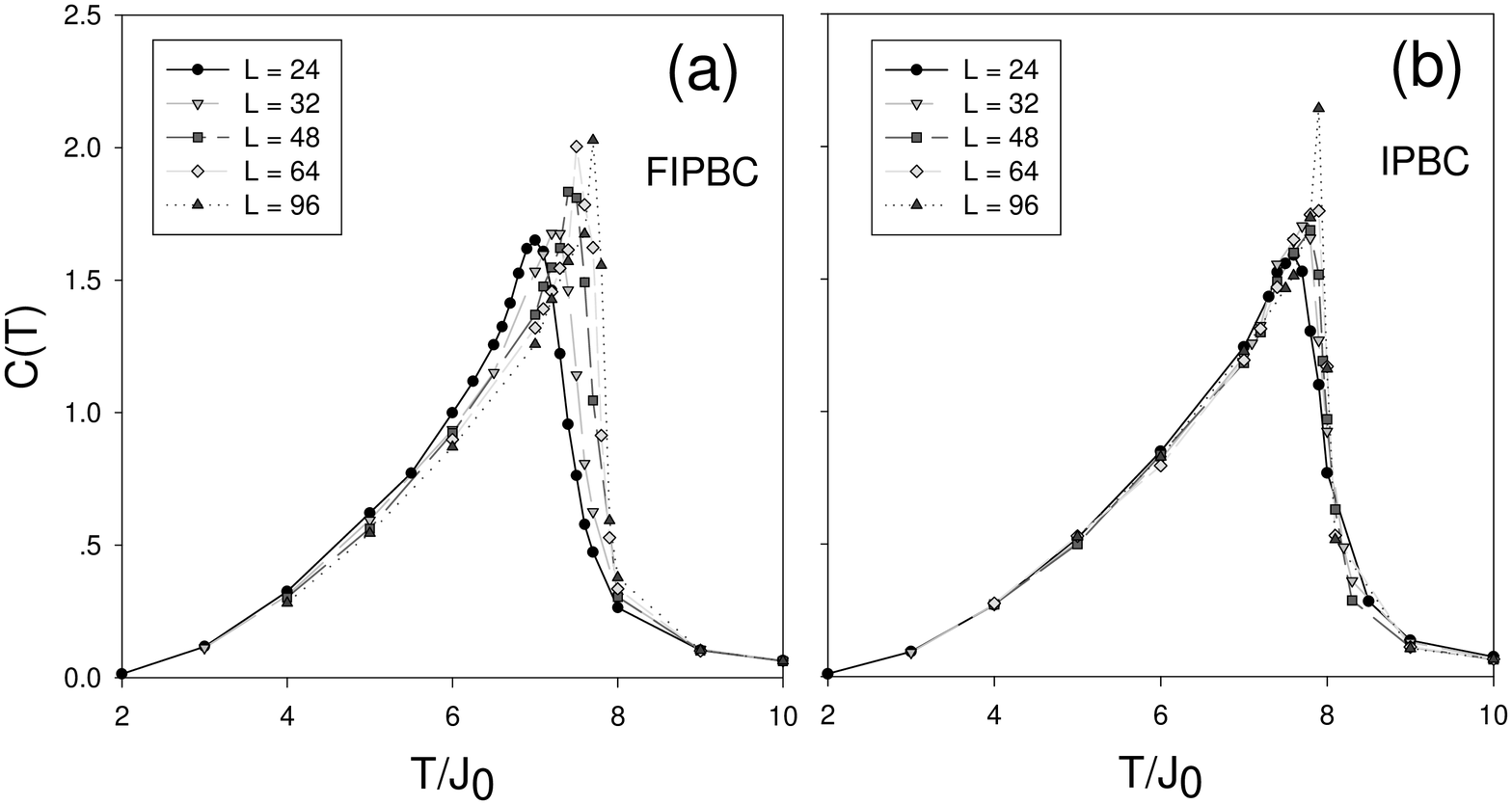}
\caption{Specific heat curves for the two-dimensional ferromagnetic model with $\alpha=3$ obtained with (a) FIPBC and (b) IPBC for different values of $L$.}
\label{fig3} 
\end{center}
\end{figure}

\begin{figure}
\begin{center}
\includegraphics[width=14cm,height=10cm,angle=0]{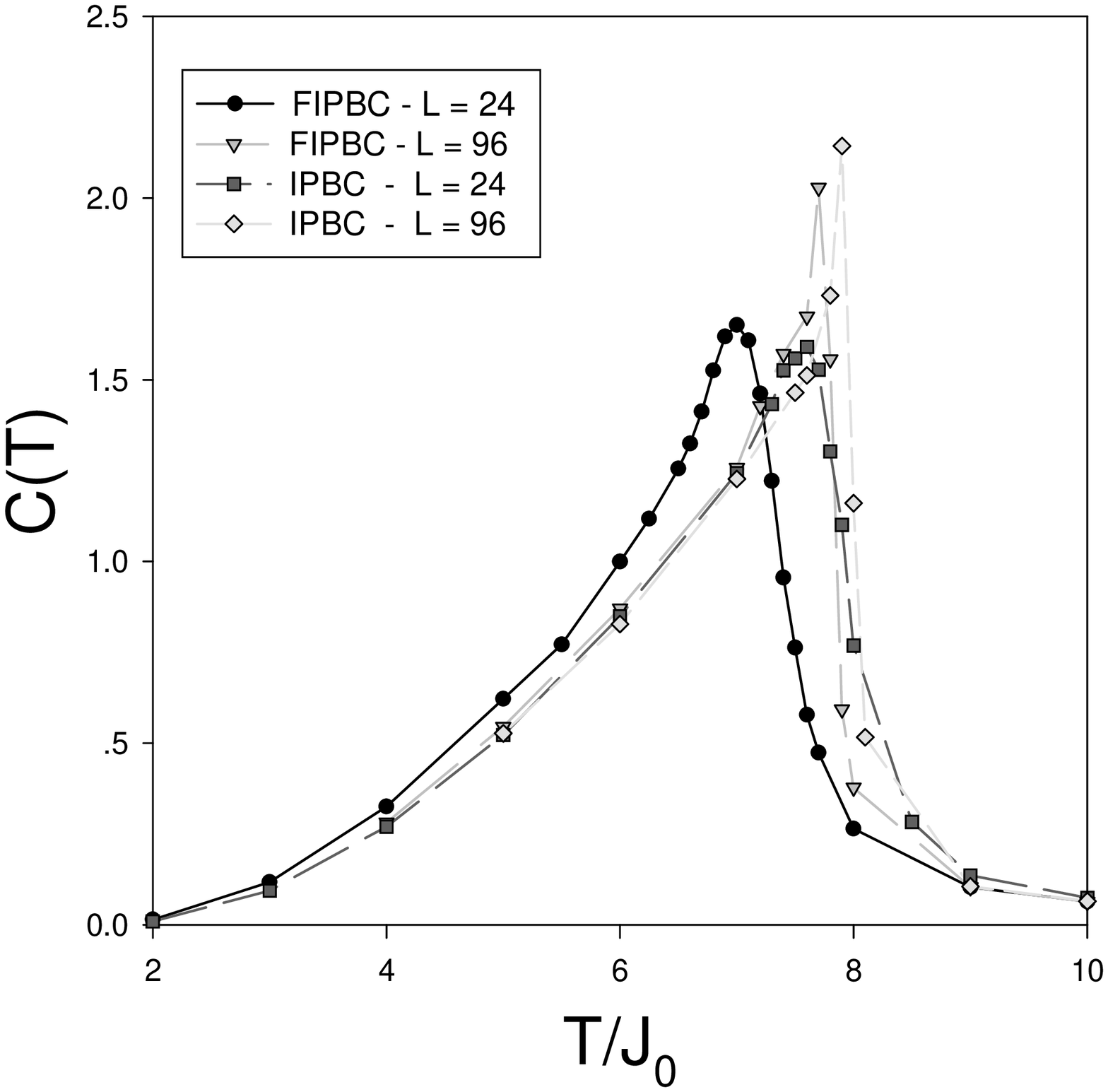}
\caption{Comparison between the specific heat curves for the two-dimensional ferromagnetic model with $\alpha=3$ obtained with FIPBC and IPBC.}
\label{fig4} 
\end{center}
\end{figure}

In this section we considered Ising models in a square lattice of $L^2$ sites with  interactions that decay as $1/r^3$. IPBC were implemented in this case using the Ewald sums technique\cite{DeBell}. We started with a Hamiltonian of the type (\ref{H1}) with $\alpha=3$ and $J_0>0$. The choice of the vale $\alpha=3$ obeys two reasons: first, it allows the usage of the Ewald sums for the implementation of the IPBC and second, because of its close relationship with dipolar interactions.
To perform the comparison between boundary conditions we calculated in this case the  specific heat

\[ C(T)= \frac{1}{NT^2} \left( \left< H^2 \right>_L-\left< H \right>_L^2\right) \]

\noindent as a function of $T$ for different system sizes $L$. In Fig.\ref{fig3} we show the specific heat calculations for both types of boundary conditions for system sizes $L=24,32,48,64$ and $96$. Although FIPBC display larger finite size effects for small system sizes than the IPBC, both type of boundary conditions give similar results when $L=96$, as can be appreciated in Fig.\ref{fig4}, where we compare the results of FIPBC and IPBC for the smallest and the largest sytem sizes. For $L=96$  the curves obtained wiht both kind of boundary conditions show clearly the emergency of a discontinuous shape, consistently with the expected classical behaviour\cite{Luijten}.

\begin{figure}
\begin{center}
\includegraphics[width=14cm,height=10cm,angle=0]{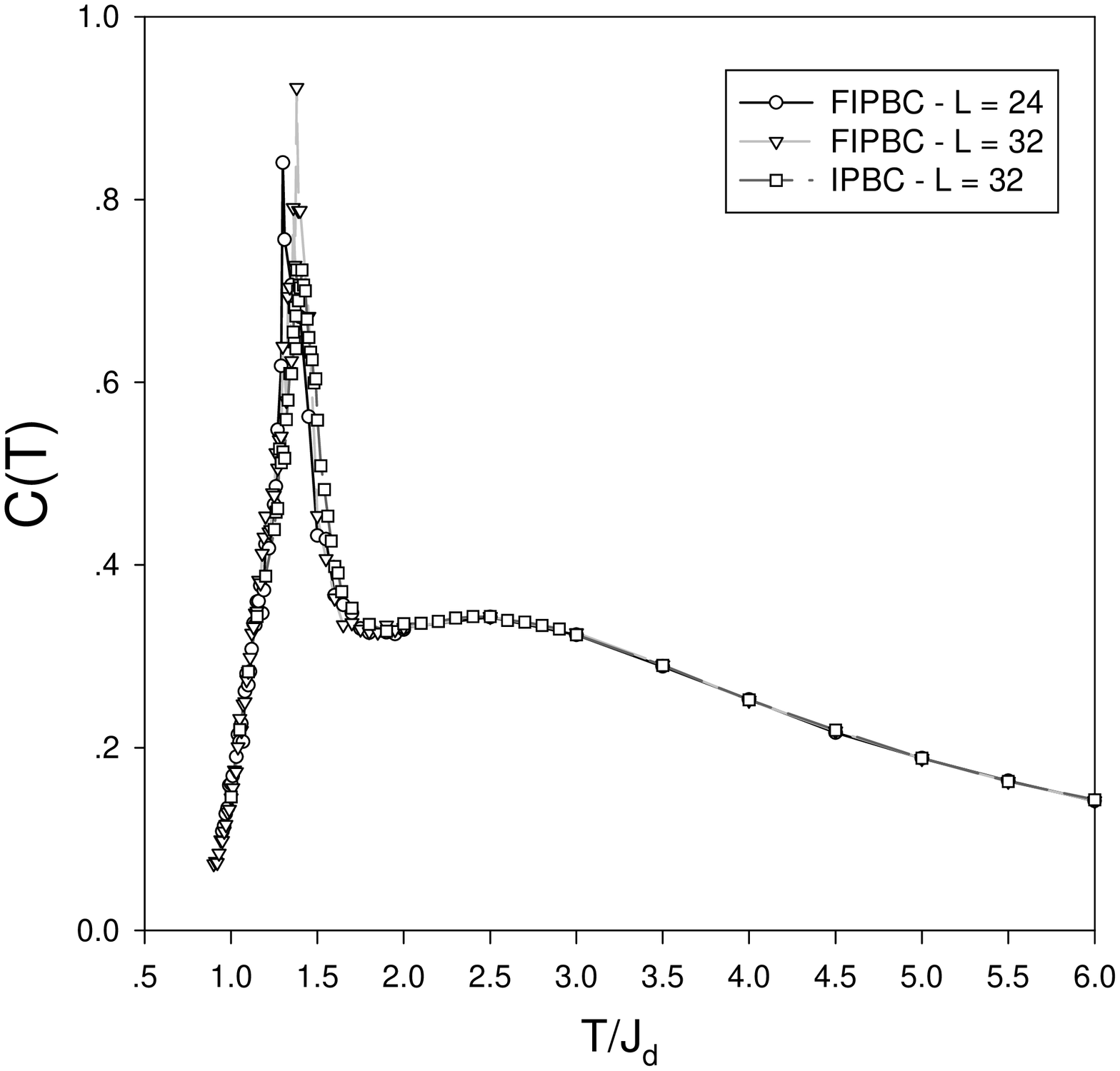}
\caption{ Specific heat vs. $T/J_d$  in the two dimensional model with competing interactions (\ref{H2}) with $\delta=J_0/J_d=3$.}
\label{fig5}
\end{center}
\end{figure}

Finally we considered the two-dimensional version of Hamiltonian (\ref{H2}). This model represents an ultrathin magnetic film where the axis of easy magnetization is oriented perpendicular to the plane of the film\cite{DeBell}. In this case the first term in Hamiltonian (\ref{H2})  represents ferromagnetic exchange interactions while the second represents dipolar interactions. The system presents a low temperature ordered phase below some critical temperature $T_c=T_c(\delta)$, composed by ferromagnetic stripes of width $h=h(\delta)$, so that spins belonging to adyacent stripes are antialigned. The overall known features of the finite temperature phase diagram of this model are described in Refs.\cite{MacIsaac}-\cite{Gleiser}. For the comparison of the boundary conditions we choose  the value $\delta=3$, for which the low temperature phase corresponds to an $h=4$ striped phase, since it is representative of the global behaviour for intermediate values of $\delta$ and was previously analized in detail by Booth et al\cite{Booth} using Monte Carlo with IPBC\cite{note}. The results are shown in Fig.\ref{fig5} for $L=24,32$. We see that   both types of boundary conditions give almost the same results even for rather small sizes, except very near the critical point (left peak; see Ref.\cite{Booth} for details) where the FIPBC give a little larger peak than the IPBC. 

\begin{table}
\begin{tabular}{|l|l|l|l|}
\hline
$\delta$ &$\alpha$ & $T_c/J_d$  (FIPBC)  & $T_c/J_d$ (IPBC)  \\
\hline   
0.5 &1.2  & $0.076 \pm 0.002$ & $0.075 \pm 0.002$\\
\hline
0.5 &1.4  & $0.060 \pm 0.002$  & $0.0625 \pm 0.002$ \\
\hline
0.5 &1.6  &$0.0575 \pm 0.002$  & $0.060 \pm 0.002$ \\
\hline
0.5 &1.8  & $0.0625 \pm 0.002$  & $0.0675 \pm 0.002$  \\
\hline   
\hline
0.9 &1.2  & $0.0575 \pm 0.002$ & $0.058 \pm 0.002$\\
\hline
0.9 &1.4  & $0.0625 \pm 0.002$  & $0.060 \pm 0.002$ \\
\hline
0.9 &1.6  &$0.061 \pm 0.002$  & $0.060 \pm 0.002$ \\
\hline
0.9 &1.8  & $0.065 \pm 0.002$  & $0.065 \pm 0.002$  \\
\hline
\hline
1.3 &1.2  & $0.077 \pm 0.002$ & $0.075 \pm 0.002$\\
\hline
1.3 &1.4  & $0.073 \pm 0.002$  & $0.073 \pm 0.002$ \\
\hline
1.3 &1.6  &$0.074 \pm 0.002$  & $0.073 \pm 0.002$ \\
\hline
1.3 &1.8  & $0.0725 \pm 0.002$  & $0.073 \pm 0.002$  \\
\hline
\end{tabular}  
\caption{\label{table2} Critical temperature $T_c/J_d$ for the one-dimensional  model with competitive interactions (\ref{H2}) obtained with FIPBC and IPBC for different values of $\alpha$ and $\delta$. }
\end{table}

\section{Discussion}

\label{conclu}

We have compared the finite size effects introduced by FIPBC and IPBC in one and two dimensional Ising models with different types of long range interactions.  This allowed us to analize a large variety of situations of interest and the influence of different parameters. We have shown that, when the interactions are such that the system presents a low temperature ordered phase with zero global magnetization the finite size effects produced by both types of boundary conditions are almost the same. This can be understood if we consider that, in this case,  the infinite sums in the local fields calculations converge fast and hence the finite size effects associated with the long range character of the interactions are wiped out quickly; hence, the truncation associated with the FIPBC  is enough to take it into account. A similar effect was observed  by Kretschmer and Binder\cite{Kretschmer} in the antiferromagnetic phase of the three dimensional Ising model with nearest neighbor exchange interactions and long range dipolar interactions. On the other hand, when the system presents low temperature ferromagnetic ordering, large differences in the finite size effects produced by the different periodic boundary conditions can be expected, depending on the values of $d$ and $\alpha$. Moreover, our results suggest that the large differences appear when the ratio $\alpha/d$ is such that the system enters into the classical regime $\alpha/d \leq 3/2$ ($\sigma<d/2$), while for $\alpha/d > 3/2$ both type of boundary conditions produce similar finite size effects, with a decreasing difference as $\alpha/d$ increases. This is also consistent with Kretschmer and Binder result \cite{Kretschmer} for $d=3$, $\alpha=3$.

 We have also shown that in the ferromagnetic one dimensional model the finite size effects given by the  FIPBC becomes very large when $\alpha\rightarrow 1^+$. This suggests that the same problem could be observed in the non-extensive regime $0\leq\alpha\leq 1$, where the FIPBC are the most direct form of implementing periodic boundary conditions. Moreover, in this regime when $\alpha<d$ the implementation of the IPBC is more involved because the infinite sums of the type (\ref{WI}) do not converge. Since in the $\alpha=0$ case  we lost the boundary completely (this corresponds, with an appropriated re-scaling,  to the Curie-Weiss model) we can expect that  in the limit $\alpha\rightarrow 0$ the influence of periodic boundary conditions will be also negligible. Hence, we may expect the largest finite size effects when using FIPBC in the region $\alpha/d \sim 1$.

Finally, as a by-product of this work, we introduced a simple method for implementing the IPBC in one dimensional models for arbitrary values of $\alpha>1$.

This work was partially supported by grants from
Consejo Nacional de Investigaciones Cient\'\i ficas y T\'ecnicas CONICET
(Argentina), Agencia C\'ordoba Ciencia (C\'ordoba, Argentina),  Secretar\'{\i}a de Ciencia y
Tecnolog\'{\i}a de la Universidad Nacional de C\'ordoba (Argentina), Funda\c{c}\~ao VITAE (Brazil) and CNPq (Brazil).  

We wish to thank R. Díaz-Méndez and Roberto Mulet for their help in the implementation of the
Ewald sums.


\begin{thebibliography}{99}
\bibitem{note} It is worth noting
that Booth and coauthors\cite{Booth} definition of the Hamiltonian
(Eq.(1) in that reference) is slightly different from ours
(Eq.(\ref{H2})).While in that paper the dipolar term contains a sum over
all pairs of spins, in Hamiltonian (\ref{H2}) we consider the sum over
every pair of spins just once. This leads to the equivalence $\delta=J/2$, $J$
being the exchange parameter in the above reference. Since the dipolar
parameter also fixes the energy units in our work, there is also a factor
$1/2$ between the critical temperatures obtained in both works.
\end{thebibliography}
\end{document}